\documentclass[reprint,preprintnumbers,nofootinbib,amsmath,amssymb,aps,pra]{revtex4-2}
\pdfoutput=1
\usepackage{graphicx,hyperref,braket,orcidlink, algorithm2e, amssymb, amsmath, dsfont}
\usepackage{lineno}
\graphicspath{{figs/}}
\hypersetup{
	bookmarks=true,
	unicode=true,
	pdftoolbar=true,
	pdfmenubar=true,
	pdffitwindow=false,
	pdfstartview={FitH},
	pdftitle={Exploring thermal equilibria of the Fermi-Hubbard model with variational quantum algorithms},
	pdfauthor={J.Y.~Araz, M.~Spannowsky, M.~Wingate}, 
	pdfsubject={Exploring thermal equilibria of the Fermi-Hubbard model with variational quantum algorithms},
	pdfcreator={J.Y.~Araz, M.~Spannowsky, M.~Wingate},
	pdfproducer={}, 
	pdfkeywords={},
	pdfnewwindow=true,
	colorlinks=true,
	linkcolor=blue,
	citecolor=magenta,
	filecolor=magenta,
	urlcolor=cyan
}

\newcommand{\tr}[1]{\mathrm{Tr}\left[#1\right]} % Trace
\begin{document}

\title{Exploring thermal equilibria of the Fermi--Hubbard model\\with variational quantum algorithms}

\author{Jack Y.\ Araz\orcidlink{0000-0001-8721-8042}}
\email{jackaraz@jlab.org}
\affiliation{Jefferson Lab, Newport News, VA 23606, United States}

\author{Michael Spannowsky\orcidlink{0000-0002-8362-0576}}
\email{michael.spannowsky@durham.ac.uk}
\affiliation{Institute for Particle Physics Phenomenology, Durham University, Durham, United Kingdom}

\author{Matthew Wingate\orcidlink{0000-0001-6568-988}}
\email{m.wingate@damtp.cam.ac.uk}
\affiliation{DAMTP, University of Cambridge, Cambridge, CB3 0WA, United Kingdom}

%\date{\today}

\preprint{IPPP/23/72, JLAB-THY-23-3968}

% \linenumbers
\begin{abstract}
This study investigates the thermal properties of the repulsive Fermi--Hubbard model with chemical potential using variational quantum algorithms, crucial in comprehending particle behaviour within lattices at high temperatures in condensed matter systems. Conventional computational methods encounter challenges, especially in managing chemical potential, prompting exploration into Hamiltonian approaches. Despite the promise of quantum algorithms, their efficacy is hampered by coherence limitations when simulating extended imaginary time evolution sequences. To overcome such constraints, this research focuses on optimizing variational quantum algorithms to probe the thermal properties of the Fermi--Hubbard model. Physics-inspired circuit designs are tailored to alleviate coherence constraints, facilitating a more comprehensive exploration of materials at elevated temperatures. Our study demonstrates the potential of variational algorithms in simulating the thermal properties of the Fermi--Hubbard model while acknowledging limitations stemming from error sources in quantum devices and encountering barren plateaus.
\end{abstract}
\maketitle

\section*{Introduction}

Quantum physics presents profound challenges in unravelling the intricate dynamics governing particle interactions within condensed matter systems. Among the pivotal models illuminating such phenomena, the Fermi--Hubbard model \cite{hubbard1963proceedings,anderson1963theory} encapsulates correlated particles' behaviour within a lattice framework. A crucial facet in comprehending these systems lies in probing their thermal properties.

However, the computational scrutiny of models encounters substantial hurdles, especially when traditional methodologies confront facets like the introduction of a chemical potential~\cite{_imkovic_2017} or topological terms~\cite{Araz:2022tbd}. Notably, Monte Carlo (MC) algorithms falter in handling such characteristics, compelling the exploration of Hamiltonian methodologies for their robustness in navigating these complexities.

Despite the promise inherent in Hamiltonian methods, challenges endure. Due to their Hamiltonian-based approach, Tensor Networks do not suffer from sign problems. Employing imaginary time evolution through techniques like Time Evolving Block Decimation (TEBD) allows for accurate simulation of thermal states for lattice field theories~\cite{Paeckel:2019yjf, Agasti_2020, PhysRevLett.93.207204, URBANEK2016170}. However, while potent, TEBD confronts computational bottlenecks; the escalating computational demand due to large bond dimensions associated with extensive time steps restricts the efficacy of such methods in capturing the thermal equilibria of these systems.

To address these challenges, attention has pivoted towards leveraging the capabilities of quantum computing. Quantum algorithms, immune to specific computational roadblocks faced by classical algorithms, offer a promising avenue for exploring complex condensed matter systems. Although it is possible to implement imaginary time evolution for quantum algorithms to study the thermal properties of lattice systems without suffering from large bond dimensions~\cite{Davoudi:2022uzo}, the finite coherence time in quantum devices imposes constraints on simulating prolonged time evolution sequences, necessitating alternative strategies. Such approaches often require a significant number of Troterrized time-evolution steps to converge to the state in question. Although such an approach guarantees the accuracy of the state up to a Trotterization error, it is not applicable to the near term devices.

This study focuses on employing variational quantum algorithms to investigate the thermal characteristics of the spin-$\tfrac12$ Fermi--Hubbard model. These techniques find their roots in the Variational Quantum Eigensolver (VQE)~\cite{Peruzzo:2014aa, Tilly:2021jem, Cerezo:2021aa}, which utilizes a trainable unitary gate sequence to approximate the diagonalization of a given Hamiltonian. Previous studies have applied this method to analyze the ground state properties of the Fermi--Hubbard model~\cite{Stanisic:2021irm, Consiglio:2021upq}. Furthermore, Refs.~\cite{Verdon:2019, Guo:2021qdb,Fromm:2023npm} have extended VQE techniques to learn thermal state density matrices using hybrid classical-quantum network systems. Despite their success, integrating classical and quantum computing methods creates unavoidable bottlenecks, as evidenced in data analysis applications~\cite{Araz:2022zxk}. It has been shown that hybrid approaches lead to longer convergence time and they require classically constructing part of the problem which will require an exponential growth in the computational resources with respect to number of qubits. These challenges have prompted exploring and developing alternative quantum-based algorithms~\cite{Foldager:2022aa, Selisko:2022wlc, Warren:2022evv, Zhang_2021}.

We employ the quantum Variational Quantum Thermaliser (qVQT) algorithm, proposed in Ref.~\cite{Selisko:2022wlc}. Through efficient circuit designs, we aim to surmount the limitations of quantum coherence constraints. By doing so, we aim to glean more profound insights into the behaviour of these materials at elevated temperatures, thereby contributing to the foundational understanding of quantum condensed matter systems. Ref.~\cite{Selisko:2022wlc} explores thermal behaviour of Heisenberg model which does not require highly complex circuit design. In this paper, we explore the limitation of qVQT algorithm for a significantly more complex quantum many-body system and show that one needs to employ more complex circuit designs in order to achieve required accuracy.

Subsequent sections delve into the methodological intricacies deployed in our study; the model will be introduced in Section~\ref{sec:hubbard}, and in Section~\ref{sec:method}, we will present our methodology, delineating the tailored circuit designs. Finally, in Sec.~\ref{sec:results}, we will discuss the insights gleaned from our investigations. This research endeavours to provide a nuanced understanding of materials at a quantum scale, potentially paving the way for advancements in material science and technological innovation.

\section{The repulsive Fermi--Hubbard model with chemical potential}\label{sec:hubbard}

The Fermi--Hubbard model belongs to the category of ``tight-binding models", representing effective electron-electron interactions by a contact term. In its second-quantized form, the Fermi--Hubbard Hamiltonian is expressed as:
\begin{eqnarray}
    \hat{H} &=& - t \sum_{\langle i,j\rangle,s}  \left[ c^{\dagger}_{i,s} c_{j,s} + {\rm h.c.} \right] \nonumber \\
     &+& U\sum_{i} \hat{n}_{i,\uparrow} \hat{n}_{i,\downarrow} -  \mu\sum_{i}\hat{n}_{i} \ .\label{eq:hubbard}
\end{eqnarray}
Here, $\hat{n}_{i,s} = c^\dagger_{i,s}c_{i,s}$ and $\hat{n}_i=\sum_s \hat{n}_{i,s}$ represent the number operator. The parameter $U$ denotes the on-site repulsive coupling constant, while $\mu$ stands for the chemical potential~\cite{hubbard1963proceedings,anderson1963theory}. The notation $\langle i,j\rangle$ denotes nearest-neighbor interaction, and the $s$ index accounts for spin orientations, where $s\in \{\uparrow, \downarrow\}$. The creation (annihilation) operators, $c_{is}^\dagger$ ($c_{is}$), follow anticommutation relations:
\begin{eqnarray}
    \left\{c_{is}, c_{js^\prime}^\dagger \right\} = \delta_{ij}\delta_{ss^\prime}\ ,\ \left\{c_{is}, c_{js^\prime} \right\} = \left\{c_{is}^\dagger, c_{js^\prime}^\dagger \right\} = 0\ . \nonumber
\end{eqnarray}
For $\mu\neq0$, Eq.~\eqref{eq:hubbard} suffers from the so-called sign problem due to the complex terms in the partition function, leading MC-based algorithms to become ill-posed~\cite{10.1063/1.4823061, PhysRevB.26.5033, 10.1063/1.447637}.

In order to embed such a Hamiltonian on a quantum device, one needs to go through a qubitization process via the Jordan--Wigner transformation~\cite{PhysRevLett.56.1529, PhysRevA.94.032338}
\begin{eqnarray}
    c^\dagger_j = \left[ \prod_{k=1}^{j-1}\left(-Z_k\right)\right] S^+_j\quad ,\quad S^\pm_j=\frac{1}{2}\left(X_j \pm i Y_j\right)\ ,\label{eq:jw}
    % c_j = \frac{1}{2} \left[ \prod_{k=1}^{j-1}\left(-Z_k\right)\right] \left( X_j - i Y_j\right)\ ,
\end{eqnarray}
where $X,\ Y$ and $Z$ represent Pauli operators. Notice that for nearest-neighbour interactions, the chain of $Z$ operators vanishes and only reappears at the periodic boundaries. By applying Eq.~\eqref{eq:jw} to Eq.~\eqref{eq:hubbard} one can rewrite the terms of the Hamiltonian $\hat{H} = H_0 + H_1 + H_2$ as 
\begin{eqnarray}
    -\frac{H_0}{t} &=& \sum_{\langle i,j\rangle,s}^{N-1} \left[ S^+_{i,s} S^-_{j,s} + {\rm h.c.}\right]  \nonumber\\&+& \left[\prod_{i=1}^{N-1} \left(-Z_i\right)\right] (S^+_{N,s} S^-_{0,s} + {\rm h.c.} ) \\
    H_1 &=& \frac{U}{4} \sum_{i}^N \left(Z_{i,\uparrow}Z_{i,\downarrow}+ Z_{i,\uparrow} +Z_{i,\downarrow} + \mathds{1}_i \right)\ , \\
    H_2 &=& -\frac{\mu}{2}\sum_{i,s}\left(Z_{i,s} + \mathds{1}_{i,s}\right)\ ,
\end{eqnarray}
where $N$ stands for the number of sites. Notice that in order to simulate both spin flavours, one needs to have two sets of $N$ qubits, one for spin-up and one for spin-down, which increases the number of qubits to $N_q = 2N$.

\section{Quantum Simulation}\label{sec:method}

In this section, we outline the methodology employed to investigate the thermal equilibria of the Fermi--Hubbard model at a fixed temperature using variational quantum algorithms. Our primary focus will be on the method illustrated in Ref.~\cite{Selisko:2022wlc}. This exercise represents the thermal mixed state induced by the Hamiltonian within quantum circuits.

The density matrix of a given Hamiltonian $\hat{H}$ at a fixed temperature can be written as:
\begin{eqnarray}
    \rho_\beta = \frac{e^{-\beta \hat{H}}}{Z}\quad , \quad Z = \tr{e^{-\beta \hat{H}}}\ ,
\end{eqnarray}
where $\beta$ is the inverse temperature ($1/T$), and $Z$ is the partition function. For a variational ansatz, the thermal state at a fixed temperature can be found by optimising the system with respect to the Helmholtz free energy $F = E - TS$, where $E$ corresponds to the expectation value of the Hamiltonian at a fixed temperature $\langle \hat{H} \rangle_\beta = \tr{\rho_\beta \hat{H}}$, and entropy, $S$, can be written as $S = -\tr{\rho_\beta\log\rho_\beta}$. The challenging part of the quantum simulation arises from the fact that a quantum computer is a pure-state simulator, and one does not have direct access to the density matrix within the circuit.

Our approach's key components involve utilising two distinct quantum circuits, each designed to address specific aspects of the problem. Our procedure is summarised in the following three subsections.

\subsection{Entropy estimation}

The first variational quantum circuit (VQC$_1$) determines the probabilities associated with each pure state within the ensemble under investigation. This circuit calculates the probability of finding the system in each pure state. Since it is not possible to extract the density matrix directly to compute the entropy, one can compute the Shannon entropy after measurement:
\begin{eqnarray}
    S = -\sum_i p_i(\theta_1)\log p_i(\theta_1)\ , \label{eq:shannon}
\end{eqnarray}
where $p_i(\theta_1) = \left\lVert\langle 0 | \hat{U}_1(\theta_1) | \phi_i\rangle \right\rVert^2$ corresponds to the probability of the $i$-th pure state, $|\phi_i\rangle$, within the ensemble. Here, $\theta_1$ represents the variational parameters of the variational quantum circuit represented by $\hat{U}_1(\theta)$.

\subsection{Estimation of the expectation value of the Hamiltonian}

The second quantum circuit (VQC$_2$) is responsible for computing the expectation value of the Hamiltonian for each of the pure states identified in the previous step. The Hamiltonian operator encapsulates the system's energy characteristics, and its expectation value reveals the energy associated with each pure state. The expectation value of the density matrix can be computed as:
\begin{eqnarray}
    \langle \hat{H} \rangle_\beta = \sum_i p_i(\theta_1) \langle \phi_i | \hat{U}_2(\theta_2) \hat{H} \hat{U}_2^\dagger(\theta_2) | \phi_i \rangle, \label{eq:expval}
\end{eqnarray}
where $\theta_2$ represents the variational parameters of the variational quantum circuit represented by $\hat{U}_2(\theta)$.

\subsection{Objective function and optimization}

With the probabilities and energies of pure states at hand, we compute the system's total free energy. This is achieved through a combination of the Shannon entropy, derived from the probabilities obtained from the first quantum circuit, and the energy obtained from the second quantum circuit. The free energy is a fundamental thermodynamic quantity that encapsulates the system's equilibrium properties.

Having established the total free energy of the system, we employ variational quantum algorithms to optimize our quantum circuits. The objective is to minimize the free energy:
\begin{eqnarray}
    \min_{\theta_{1,2}} F(\theta_1,\theta_2)\ , \nonumber
\end{eqnarray}
which corresponds to finding the variational parameters that best represent the system's thermal equilibrium. The variational parameters, $\theta_{1,2}$, can be updated with respect to the gradient of the free energy:
\begin{eqnarray}
    \theta^\prime_2 = f\left(\theta_2, \frac{\partial \langle \hat{H} \rangle_\beta}{\partial \theta_2}\right)\quad , \quad \theta^\prime_1 = f\left(\theta_1, \frac{\partial \langle \hat{H} \rangle_\beta}{\partial \theta_1} - T\frac{d S}{d \theta_1}\right), \nonumber
\end{eqnarray}
where $f$ represents a function to be employed to update variational parameters, e.g.\ gradient descent. This optimization process enables us to refine our quantum states, ultimately leading to a more accurate representation of the system's behaviour.

\section{Results}\label{sec:results}

We tested the algorithm described in the previous section on a four-site system, which required eight qubits to describe the Fermi--Hubbard model, with four qubits assigned to each spin configuration.

For VQC$_1$, we selected a strongly entangling layer configuration as prescribed in Ref.~\cite{Schuld_2020}. We observed that maximising the entanglement structure within the circuit was crucial for accurately representing the entropy of the model. Other ansatze did not yield satisfactory results\footnote{We also tested our algorithm using the initially proposed circuit structure for VQC$_1$ in Ref.~\cite{Selisko:2022wlc}, which is based on simple Pauli rotations per qubit.}.

For VQC$_2$, we adopted the variational circuit structure proposed in Ref.~\cite{Consiglio:2021upq}. This configuration implements a Trotterised Hamiltonian time evolution sequence as a variational ansatz time step for each term, which is implemented as a variational parameter. One can write Eq.~\eqref{eq:hubbard} as $\hat{U}_2(\theta_2) = \left(e^{i\theta_2 \hat{H}}\right)^L$:
\begin{eqnarray}
    \hat{U}_T(\theta_2) &=& \prod_{\langle i,j\rangle, s} e^{i\left( \theta^0_{ijs} X_{i,s}\otimes X_{j,s} + \theta^1_{ijs} Y_{i,s}\otimes Y_{j,s}\right)}\ , \label{eq:ansatz_t}\\
    \hat{U}_U(\theta_2) &=& \prod_{i,s} e^{i \left( \theta^0_{is}Z_{i,s}\otimes Z_{i,s+1} + \theta^1_{is}Z_{i,s} \right) }\ , \label{eq:ansatz_u}
\end{eqnarray}
where $\{\theta^{0,1}_{ijs}, \theta^{0,1}_{is}\}\in \theta_2$ each being independent, and $X,\ Y$ and $Z$ represent the Pauli operators, and subscripts indicate the qubits that they are acting on. Notice that $\theta_2$ indicates the collective set of parameters included within VQC$_2$, each term has an independent parameter. The complete ansatz per layer consists of 
\begin{eqnarray}
    \hat{U}_2(\theta_2) = \prod^L_{k=1}\hat{U}_T(\theta^{(k)}_2) \hat{U}_U(\theta^{(k)}_2)\ ,\nonumber
\end{eqnarray}
where $L$ refers to the number of independent layers. The difference between Eqs.~\eqref{eq:ansatz_t}\footnote{Also known as Fermionic SWAP gate, see Ref.~\cite{Hashim:2021loh}.} and~\eqref{eq:ansatz_u} are coming from hopping terms and repulsive term in the Hubbard model respectively. Here, subindices $i,\ j$ correspond to the site index, $s$ to the spin index and $k$ is the layer index. This step ensures the preservation of the physical properties of spin types and the correlation between spin flavours induced by on-site repulsive coupling. We refer the reader to Appendix~\ref{app:ansatz} for details on the circuit representation.

For system optimisation, we used SciPy's minimiser based on Sequential Least Squares Programming~\cite{2020SciPy-NMeth} (version 1.10.0). The quantum circuit and its gradient were simulated using PennyLane~\cite{bergholm2020pennylane} (version 0.32.0) and Qiskit's AER simulator~\cite{Qiskit} (\texttt{qiskit.aer} plugin version 0.28.0, Qiskit version 0.24.1). All parameters are initialised from a central Gaussian distribution with 1 standard deviation. To evaluate the quality of the final density matrix, we computed the fidelity of the reconstructed density matrix with respect to the one computed via exact diagonalisation:
\begin{eqnarray}
    \text{Fidelity}(\rho, \rho_{\text{rec}}) = \left( \tr{\sqrt{\sqrt{\rho_{\text{rec}}}\rho\sqrt{\rho_{\text{rec}}}}}\right)^2 \ , \nonumber
\end{eqnarray}
where $\rho_{\text{rec}}$ represents the reconstructed density matrix.
\begin{figure}[!h]
    \centering
    \includegraphics[width=\linewidth]{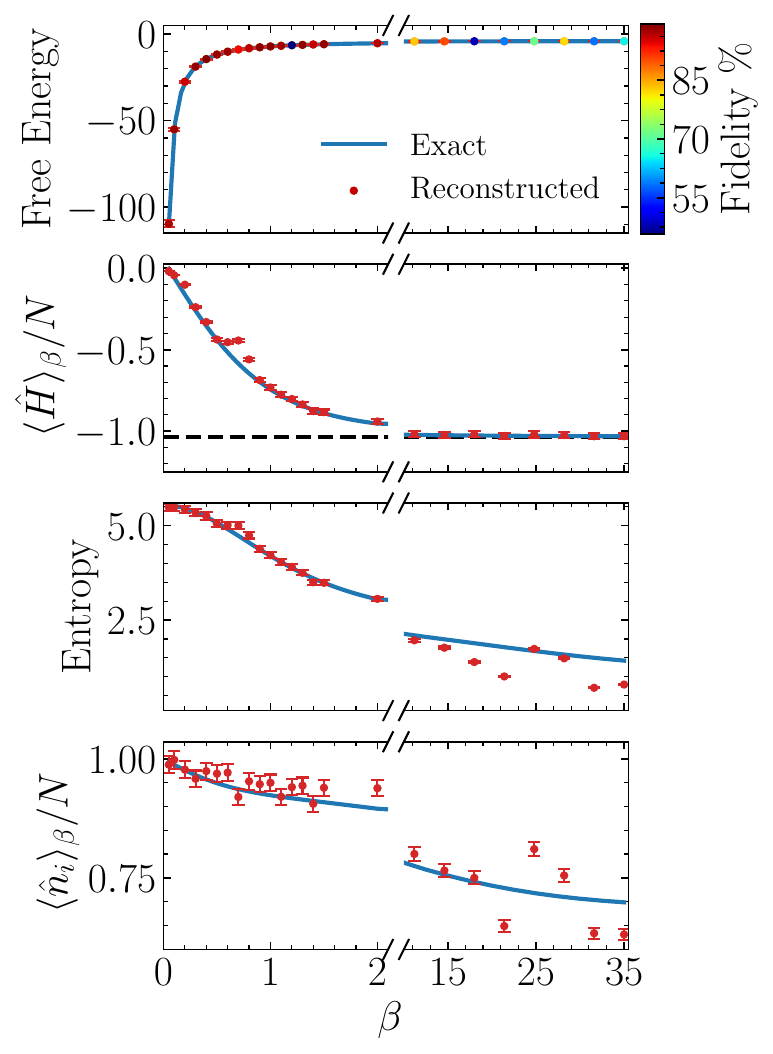}
    \caption{\it Quality metrics computed for fixed $U=0.8$ and $\mu=0.2$ values for four-site Fermi--Hubbard model. The blue line in each panel shows the results computed via exact diagonalisation, where dots represent the results reconstructed via quantum simulation. The coloured dots in the top panel show the Fidelity value achieved for each point. From top to bottom, the panels compare free energy, Hamiltonian density, entropy and number operator density. The error bars represent the statistical uncertainty for 3000 shots, and the black dashed line shows the expectation value of the operators at $\beta\to\infty$.}
    \label{fig:fe_energy_entropy_number_4site}
\end{figure}

% To study the impact of temperature on the results, we scanned the inverse temperature $\beta\in [0.05, 35]$ while fixing $U=0.8$ and $\mu=0.2$. Given that the VQC prescription above represents a single layer, we applied four of these layers for each VQC as an initial ansatz. Fig.~\ref{fig:fe_energy_entropy_number_4site} displays the results of this scan for five different metrics. The top panel shows the free energy of the system computed via exact diagonalisation (blue line) and the reconstructed free energy as a result of optimisation (coloured dots). The colour of the dots in this plot represents the fidelity between the exact and reconstructed density matrix.%, and for this scan, we did not observe any value below 93\%.

% We observed a minimum of 93\% fidelity within $\beta\in [0.05, 2]$ range where our initial four layer proposal was quite successful. However, for larger $\beta$ values, despite reconstructing the free energy accurately, fidelity has been observed to be significantly low. This is due to the reduced importance of the entropy term in the free energy. As $\beta$ increases, system has been observed to have harder time to converge into the correct entropy value. It is essential to note that although some values were spot on, this has been observed to be not by achieving physical equilibrium. In order to estimate the physical accuracy of the system independently, we used the number operator density, which has been shown to deviate from its exact value quite significantly in higher $\beta$ values.

To investigate the influence of temperature on the outcomes, we conducted a scan across the inverse temperature range $\beta\in [0.05, 35]$ while maintaining fixed values for $U=0.8$ and $\mu=0.2$. The VQC prescription mentioned earlier represents a single layer, and we implemented four such layers for each VQC as an initial ansatz. The results of this scan for four different metrics are illustrated in Fig.~\ref{fig:fe_energy_entropy_number_4site}.

In the top panel of the figure, the computed free energy of the system via exact diagonalisation is depicted by the blue line, while coloured dots represent the optimised reconstructed free energy. The colour of these dots reflects the fidelity between the exact and reconstructed density matrix.

Within the range $\beta\in [0.05, 2]$, we observed a minimum fidelity of 93\%, indicating considerable success with our initial four-layer proposal. However, for higher $\beta$ values, despite accurately reconstructing the free energy, we noticed a significant decrease in fidelity. This decline is attributed to the reduced significance of the entropy term in the free energy as $\beta$ increases. Consequently, the system encountered greater difficulty converging to the correct entropy value with increasing $\beta$. Notably, while some values matched well, they did not reflect achieving physical equilibrium. In order to assess the importance of such large $\beta$ values, we also include the expectation value of the operators at $T=0$, shown via a black dashed line. This indicates the system behaves like a ground state at $\beta\gtrsim10$.

To independently assess the physical accuracy of the system, we utilised the number operator density, $\sum_i\hat{n}_i$, revealing considerable deviations from its exact value at higher $\beta$ values. This highlights the necessity of considering physical equilibrium beyond just achieving specific numerical accuracies. The results within $\beta>10$ region have been constructed by increasing the number of layers of the ansatz. However, we did not observe significant improvement in the fidelity for up to five layers.

\begin{figure*}[t]
    \centering
    \includegraphics[scale=0.6]{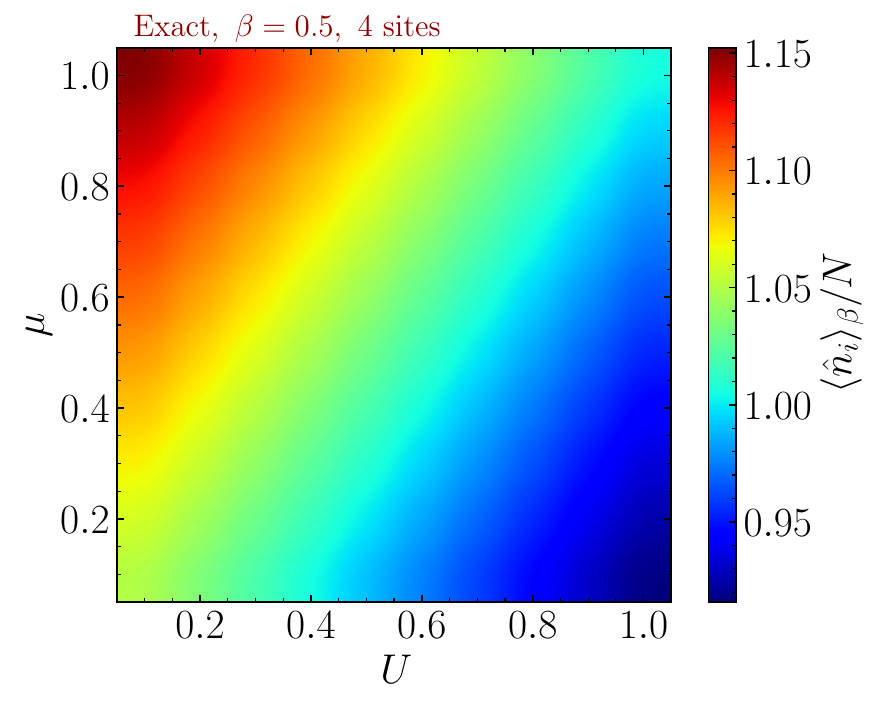}
    \includegraphics[scale=0.6]{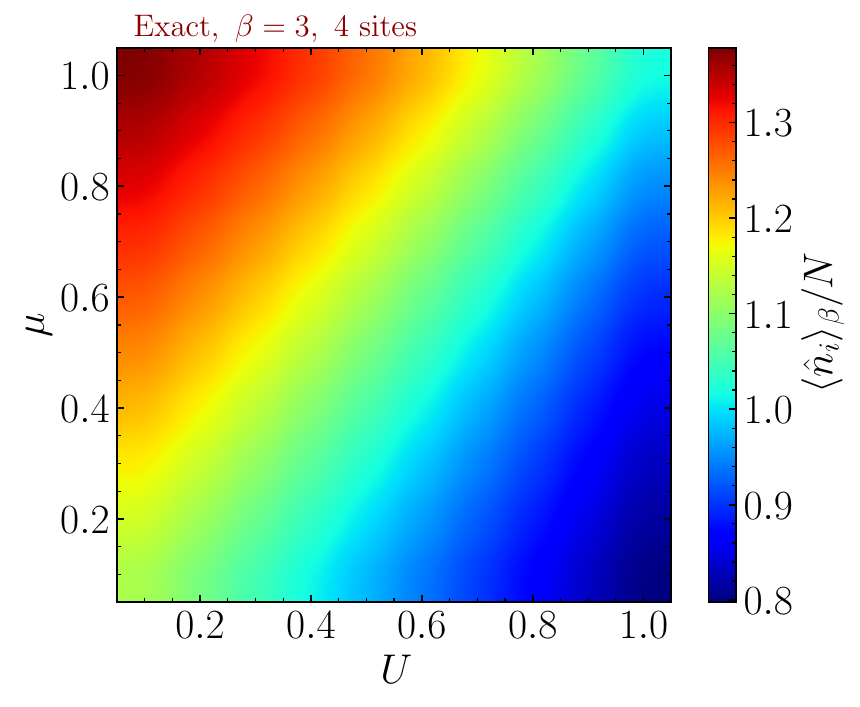}\\
    \includegraphics[scale=0.6]{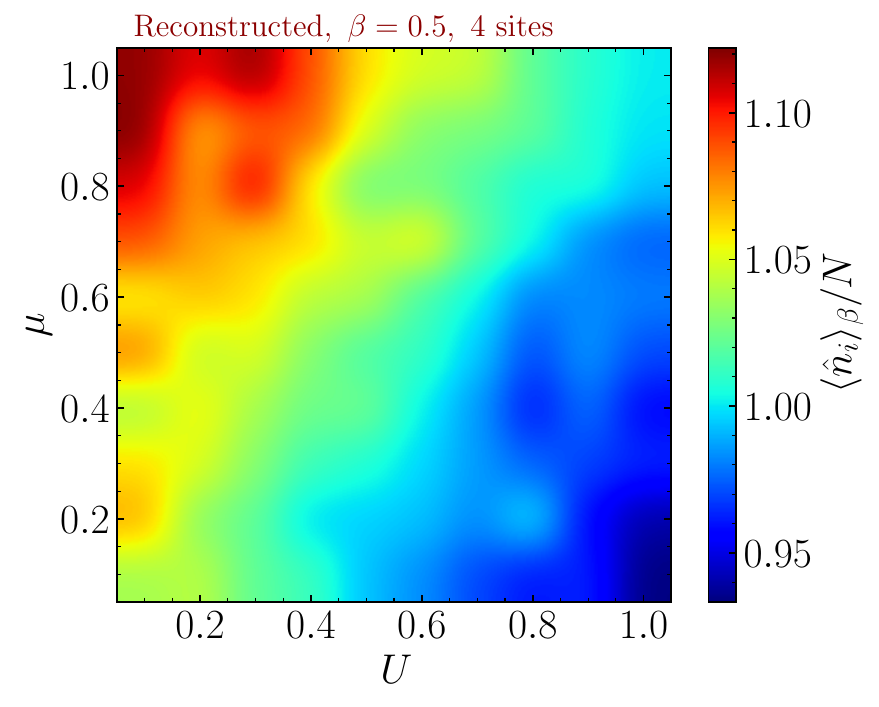}
    \includegraphics[scale=0.6]{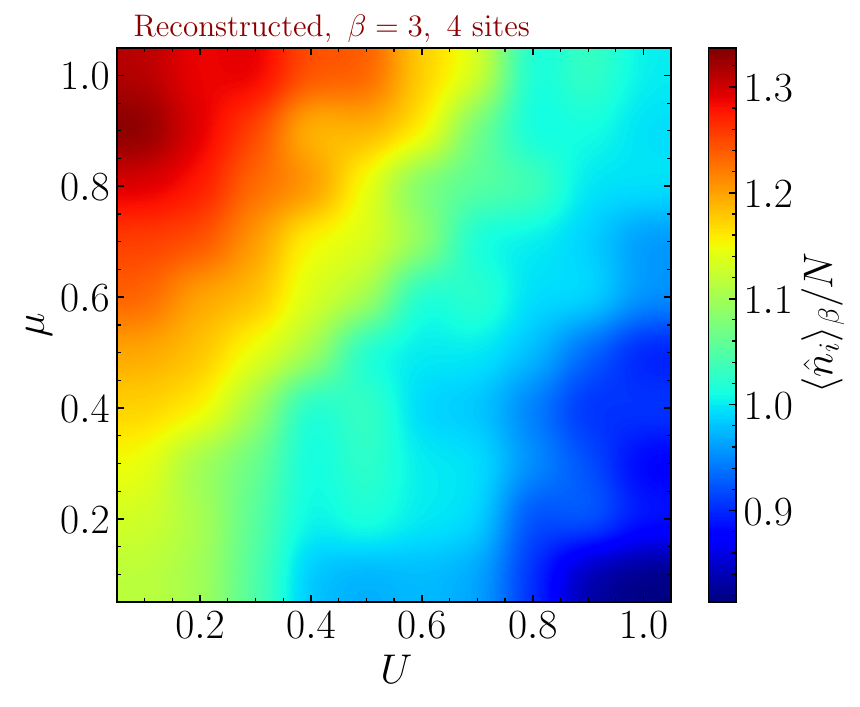}
    \caption{\it The top panels show the results computed with exact diagonalisation, whereas the bottom panels show the same results reconstructed with the variational quantum algorithm at a fixed temperature, $T=2$ on the left and $T=0.\bar{3}$ on the right. The colour in each panel shows the number operator density per $(U,\mu)$ grid.}
    \label{fig:u_mu_scan_beta}
\end{figure*}

In order to assess the possible effects on more generic scans, we computed the density of the number operator at a fixed temperature for $U,\ \mu\in[0.1,1]$.  Figure~\ref{fig:u_mu_scan_beta} shows the number density value (represented by the change in colour) for fixed $\beta$ values, 0.5 and 3, respectively. The top panel in each figure shows the result from exact diagonalization, whereas the bottom panel shows the results reconstructed by quantum simulation. We observe reasonable results for both $\beta$ values, matching the results from exact diagonalization.

\subsection{Cost estimation}

The variational system under consideration exhibits two sources of computational complexity. The first arises from computational errors within the quantum device, while the second stems from the optimization landscape. We quantified the CNOT gates within each quantum circuit to estimate the error source. However, it is important to note that the precision required for implementing rotation gates would amplify the number of T gate implementations within the circuit, thereby introducing an additional source of error. For simplicity, we focused solely on the CNOT count.

The generalized CNOT count for both circuits in terms of the number of sites ($N$) and layers ($L$) can be expressed as follows:
\begin{eqnarray}
    \#{\rm CNOT}_{\rm VQC_1} &=& 2LN\ , \label{eq:complexity1}\\
    \#{\rm CNOT}_{\rm VQC_2} &=& 2L (5N - 4)\ . \label{eq:complexity2}
\end{eqnarray}
These equations demonstrate that both circuits scale the CNOT count linearly concerning both $N$ and $L$. Here, Eq.~\eqref{eq:complexity1} denotes the CNOT count in VQC$_1$, while Eq.~\eqref{eq:complexity2} represents VQC$_2$. Additionally, fully connected quantum circuit and Hamiltonian based approach has 24 and 28 trainable parameters per layer, respectively.

\begin{figure}
    \centering
    \includegraphics[width=\linewidth]{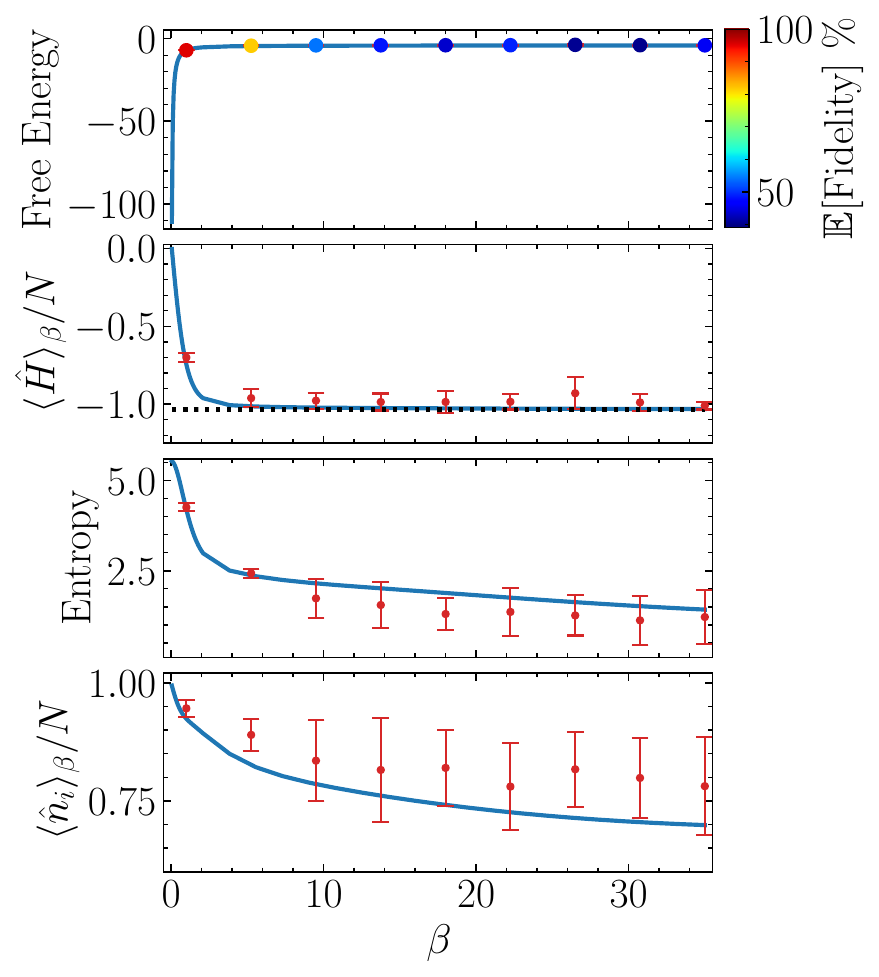}
    \caption{\it Same as Fig.~\ref{fig:fe_energy_entropy_number_4site} but error bars represent the optimisation error, estimated from 10 different initialisation. The blue line shows the results from exact diagonalisation and red dots represents the results from the mean of different initialisation between $[-\pi,\pi]$.}
    \label{fig:opterr}
\end{figure}

For variational algorithms, an additional source of error arises from optimization. Depending on the initialization, the optimization algorithm can explore different regions of the objective function landscape, which, in this specific application, represents the free energy. In order to test this we sampled our trainable parameters from a uniform distribution within $[-\pi,\pi]$. Fig.~\ref{fig:fe_energy_entropy_number_4site} displays the results for one such initialization, including shot noise. Fig.~\ref{fig:opterr} presents the same computation as Fig.~\ref{fig:fe_energy_entropy_number_4site} for 10 different initialisations. In this figure, the blue line represents the exact diagonalisation results, while the red dots denote results obtained from optimization. The error bars represent one standard deviation from the mean value of the 10 initializations. We observe significantly larger fluctuations in the computation compared to the statistical error from the shot samples. Large uncertainties indicate that the objective lanscape has many local minimum and optimisation algorithm can get stuck in these. Additionally, Table~\ref{tab:numbers} shows the number of layers that has been used in first and second circuit followed by number of iterations that it took for optimisation algorithm to converge. Note that the maximum number of layers allowed is limited by 5 and the algorithm iteratively increases the layers if fidelity is under 90\%. The respective circuit layers are increased depending on the source of discrepancy, i.e. energy or entropy. We observed that, although the convergence rate is improved with larger $\beta$ values, number of layers required in the first circuit is consistently high. This indicates that the algorithm struggles to find an optimal entropy value and since the entropy is suppressed by $\beta$ it gets harder to find an optimal entropy value at large $\beta$.

\begin{table*}
    \centering
    \begin{tabular}{l||ccccccccc}
    \hline
    $\beta$ & $1.00$ & $5.25$ & $9.50$ & $13.75$ & $18.00$ & $22.25$ & $26.50$ & $30.75$ & $35.00$ \\\hline\hline
        $\#L_{{\rm VQC}_1}$ & $3.3 \pm 0.9$ & $4.5 \pm 0.7$ & $4.3 \pm 0.6$ & $4.2 \pm 0.7$ & $4.5 \pm 0.9$ & $3.8 \pm 1.1$ & $3.6 \pm 1.2$ & $4.0 \pm 0.9$ & $3.9 \pm 1.3$\\
        $\#L_{{\rm VQC}_2}$ & $3.7 \pm 0.9$ & $2.5 \pm 0.7$ & $2.1 \pm 0.5$ & $2.4 \pm 0.5$ & $2.4 \pm 0.7$ & $2.0 \pm 0.7$ & $2.1 \pm 0.7$ & $2.1 \pm 0.8$ & $2.5 \pm 0.7$\\
        $\#{\rm Iterations}$ & $123.1 \pm 44.4$ & $116.5 \pm 36.4$ & $91.6 \pm 27.3$ & $88.0 \pm 17.1$ & $94.2 \pm 31.3$ & $97.4 \pm 25.5$ & $77.2 \pm 24.8$ & $87.0 \pm 25.1$ & $72.0 \pm 23.0$\\
        \hline
    \end{tabular}
    \caption{\it Number of layers for both circuits and optimisation steps has been used during the optimisation process. Each value is presented as the mean of 10 independent initialisation and one standard deviation from the mean.}
    \label{tab:numbers}
\end{table*}

\begin{figure}[!t]
    \centering
    \includegraphics[width=\linewidth]{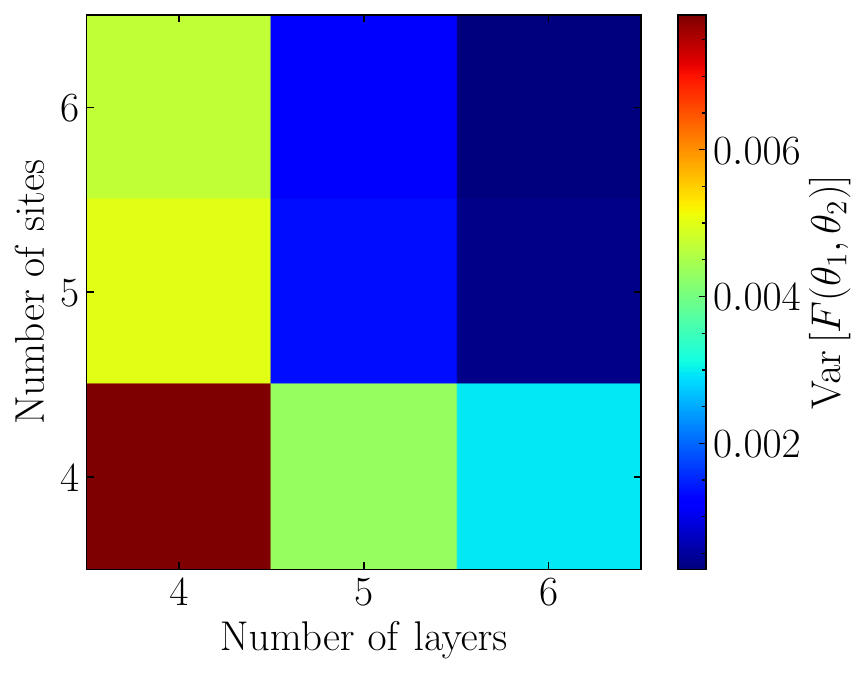}
    \caption{\it Changes in the variance of the free energy with respect to the number of sites and layers in the variational circuits where both circuits are assumed to have the same number of layers. Variance has been estimated by drawing 500 samples for $\theta_1$ and $\theta_2$.}
    \label{fig:var}
\end{figure}

Furthermore, an essential aspect of optimizing the system involves estimating the presence of barren plateaus. This was accomplished by computing the variance in the free energy based on 500 different $\theta$ values drawn from a uniform distribution with $\theta_{1,2}\in [-\pi,\pi]$.\footnote{It has been shown that the variance in gradient is equivalent to the variance in the objective function~\cite{Ragone:2023qbn}.} Fig.~\ref{fig:var} illustrates the variance change concerning the number of layers and sites, where the number of qubits is twice the number of sites. Our observations revealed that the variance in the free energy remains around $\mathcal{O}(10^{-3})$ up to 6 layers and sites without a significant decline. This suggests that the algorithm can be optimized for a relatively large number of qubits and layers. It is essential to emphasise that using ansatze, which is typically used for quantum machine learning applications, has led to significantly lower values for the variance in free energy. This finding reveals the importance of using physics-inspired ansatz to prevent or reduce the risk of barren plateaus in variational quantum simulation applications.

\section{Conclusion}

In this study, we investigated the feasibility of variational quantum algorithms in examining the thermal behaviour of the Fermi--Hubbard model. Our analysis demonstrates that employing qVQT alongside a physics-inspired ansatz proves to be a viable approximation technique. Notably, while achieving a relatively accurate depiction of the free energy value at a given temperature is possible with any ansatz, our findings underscore that the physics-inspired ansatz offers superior approximation, effectively capturing the system's physical behaviour. To gauge this, we utilized the number density operator as an independent metric, separate from the optimization process. We showed that, although higher $\beta$ values pose a significant challenge for the optimisation algorithm due to the large suppression of the entropy term, it is possible to increase the fidelity of the system by increasing the number of layers in the VQC$_1$.

However, the Fermi--Hubbard model's gapless nature necessitates a substantial number of sites to capture the lattice's phase transition behaviour. Unfortunately, our computational resources limited our ability to explore larger site numbers. Nonetheless, our estimations indicate that the proposed ansatz holds promise in providing a relatively barren plateau-free environment for significantly larger lattices. Implementing such approaches on tensor networks (TNs) demands substantial resources due to the exponential expansion in the bond dimension during the state's imaginary time evolution (e.g., TEBD algorithm). Despite the significant limitations of current quantum computers, they are not anticipated to suffer from bond dimension constraints.

\section*{Acknowledgements}

JYA acknowledges the hospitality of Cambridge University. JYA is supported by the U.S. Department of Energy, Office of Science, Contract No. DE-AC05-06OR23177, under which Jefferson Science Associates, LLC operates Jefferson Lab and in part by the DOE, Office of Science, Office of Nuclear Physics, Early Career Program under contract No DE-SC0024358. MW acknowledges support from STFC consolidated grant ST/X000664/1 and an IPPP Associateship.

\appendix
\section{Physics-inspired ansatz}\label{app:ansatz}

\begin{figure}[t]
    \centering
    \includegraphics[scale=0.46]{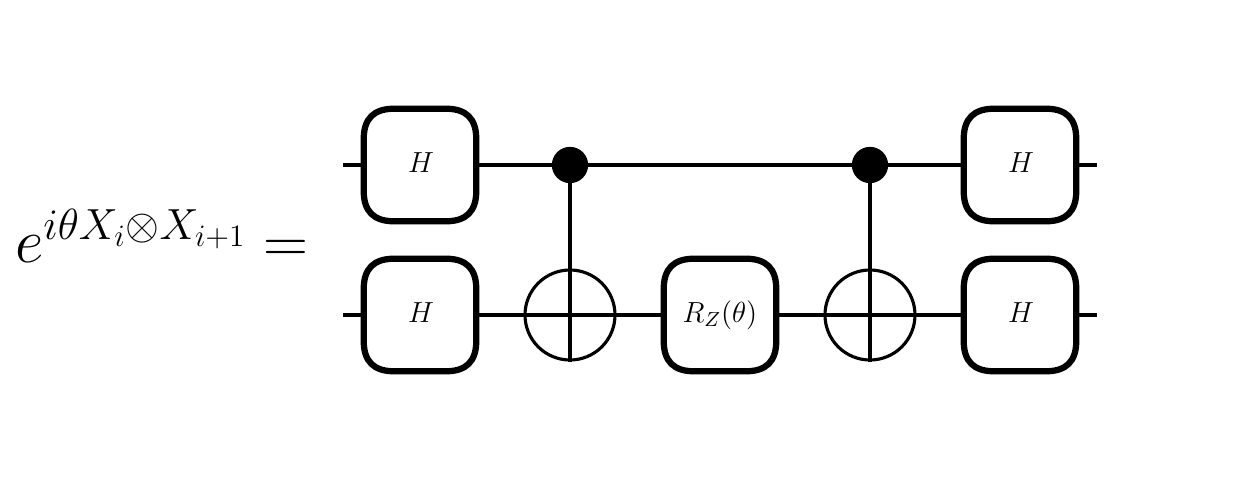}
    \includegraphics[scale=0.46]{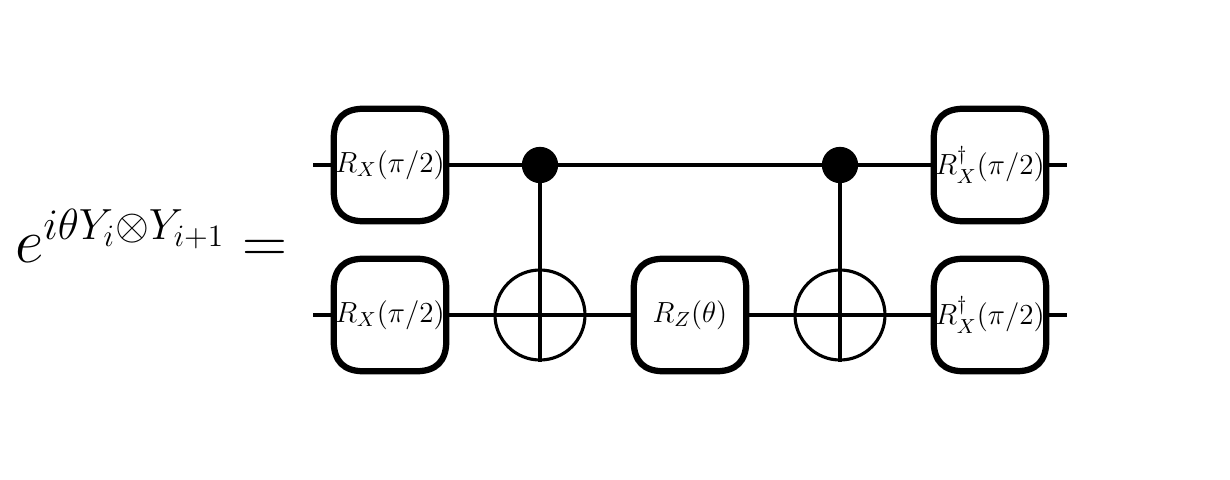}
    \includegraphics[scale=0.46]{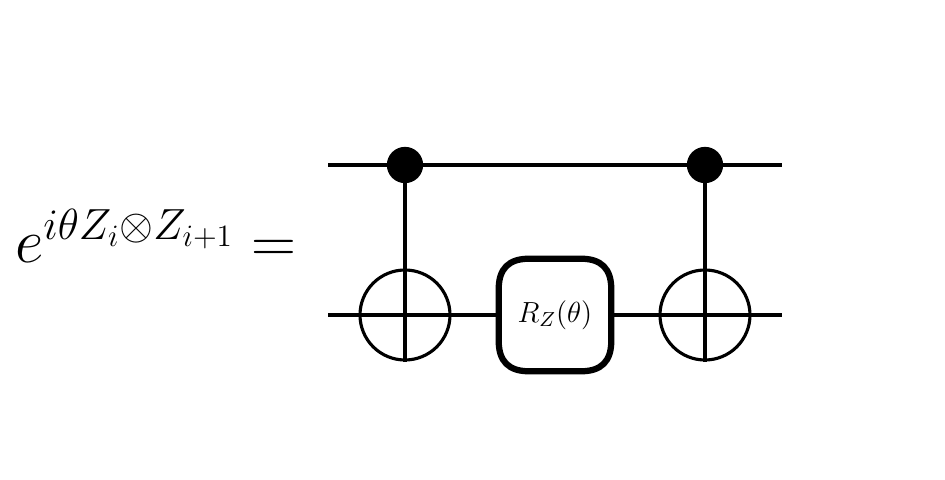}
    \caption{\it Individual terms of the physics-inspired ansatz have been represented as pieces of the quantum circuit. $\theta\in[-\pi,\pi]$ represents trainable parameter in each term and $R_m,\ m\in\{X,\ Z\}$ represents rotation in respective Pauli axis.}
    \label{fig:vqc2_ansatz}
\end{figure}

The ansatz that is proposed for VQC$_2$ in Eqs.~\eqref{eq:ansatz_t} and~\eqref{eq:ansatz_u} has been inspired by the Trotterised time evolution steps of Fermi--Hubbard model in Eq.~\eqref{eq:hubbard}. The circuit version of the three main terms has been represented in Fig.~\ref{fig:vqc2_ansatz}. These terms can be realised using similarity transformations where $Y=SHZHS^\dagger$ and $Z=HXH$. Here $H$ stands for Hadamard gate, and $S$ stands for phase gate with $\lambda=\pi/2$. Notice that $SHS$ can also be expressed as $R_X(-\pi/2)=R_X^\dagger(\pi/2)$. 

\bibliography{bibliography}
\end{document}